# Virtual Optical Pulling Force

Sergey Lepeshov[1,*], and Alex Krasnok[2,*]

[1]ITMO University, St. Petersburg 197101, Russia

[2]Photonics Initiative, Advanced Science Research Center, City University of New York, New York 10031, USA

*To whom correspondence should be addressed: s.lepeshov@gmail.com, akrasnok@gc.cuny.edu

## Abstract

The tremendous progress in light scattering engineering made it feasible to develop optical tweezers allowing capture, hold, and controllable displacement of submicronsize particles and biological structures. However, the momentum conservation law imposes a fundamental restriction on the optical pressure to be repulsive in paraxial fields. Although different approaches to get around this restriction have been proposed, they are rather sophisticated and rely on either wavefront engineering or utilize active media. Herein, we revisit the issue of optical forces by their analytic continuation to the complex frequency plane and considering their behavior in transient. We show that the exponential excitation at the complex frequency offers an intriguing ability to achieve a pulling force for a passive resonator of any shape and composition even in the paraxial approximation, the remarkable effect which is not reduced to the Fourier transform. The approach is linked to the "virtual gain" effect when an appropriate transient decay of the excitation signal makes it weaker than the outgoing signal that carries away greater energy and momentum flux density. The approach is implemented for the Fabry-Perot cavity and a high refractive index dielectric nanoparticle, a fruitful platform for intracellular spectroscopy and lab-on-a-chip technologies where the proposed technique may found unprecedented capabilities.

## Introduction

Light scattering is ubiquitous and plays an essential role in the study of nature and conquering light-matter interactions for modern technology. According to A. Einstein [1,2], the light quanta (photons) carry the energy $\hbar\omega_0$ and momentum $\hbar\mathbf{k}$, where $\omega_0$ and $\mathbf{k}=\omega/c$ are the frequency and wavenumber of a photon, $\hbar$ is the reduced Planck's constant, and $c$ is the speed of light. Hence, every act of light scattering, when a photon changes its direction (or disappears), is accompanied by a transition of a portion of its momentum to the object. If the object is light enough and the photon momentum flux ($N\hbar\mathbf{k}$, $N$ is the photon density) is large enough, this momentum transition can be detected through a mechanical motion of the object, as has been done for the first time by Lebedev [3] and Nichols [4]. In the early years, this phenomenon helped to establish the theory of light, but after the work of Arthur Ashkin [5–7] (Nobel Prize in Physics, 2018), optical forces became a powerful catalyst for modern technologies. Namely, it has been shown that optical scattering allows controlling the position of small objects using purely electromagnetic forces. In such techniques, the intensity gradient force is used to trap the object in the lateral plane, whereas the radiation pressure allows controlling the object's position along the beam. Today, this laser trapping became routine for different kinds of optical spectroscopy of single particles and living cells, optical tweezers [8–10], optical binding [11], laser cooling, lab-on-a-chip technologies to name just a few [12].

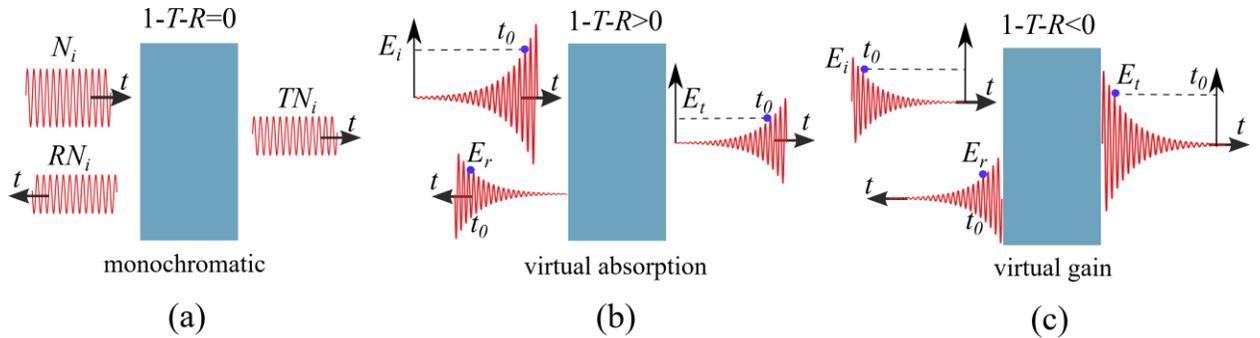

FIG. 1. (a) Schematic of light scattering from a (lossless) dielectric slab. The vector sum of all momenta gives the optical pressure $F=N_i\hbar k(1+R-T)>0$. (b) Virtual absorption. As the incoming signal grows in time exponentially faster than the decay rate of the cavity, the incoming signal $|E_i|$ can be greater than outgoing signals, $|E_t|+|E_r|$, even in the lossless case. The remaining part gets stored within the cavity (virtual loss) until the incoming signal keeps the exponential growth. In this case, $1-T-R>0$ and the optical pressure can be increased. (c) Virtual gain. As the incoming signal exponentially decays faster than the cavity decay rate, the output

signal $|E_t|+|E_r|$ can be greater than the input one, giving rise to the gain behavior. In this case, $1-T-R<0$ and the optical pressure can become negative.

In the paraxial approximation, when the size of the object is small enough, and the beam is unfocused, the intensity gradient over the object size vanishes, and the radiation pressure plays the key role. In this approximation, the momentum conservation law imposes a fundamental restriction on the optical pressure to be repulsive. Indeed, the momentum flux of the scattered light along the wave vector of incident light ($\mathbf{k}_i$) through the object's surface is smaller than the incident momentum flux, and hence the resulting radiation pressure is obliged to be repulsive. For example, consider the case of radiation pressure for a lossless dielectric slab, Figure 1(a), lit by a monochromatic incident wave with the photon density per unit square and time duration $N_i = c\varepsilon_0 E_0^2/2\hbar\omega$, where $E_0^2$ is the electric field intensity, and $\varepsilon_0$ is the dielectric constant. This incident light carries the density of momentum $N_i\hbar k = N_i\hbar\omega/c$. A part of this momentum gets transmitted through the slab ($TN_i\hbar k$), and another part gets reflected towards the source ($RN_i\hbar k$). Here $T$ and $R$ are transmission and reflection coefficients. The total pressure on the slab is the vector sum of all these momentum flux densities [13]

$$F(\omega) = N_i\hbar k\left[1+R(\omega)-T(\omega)\right]. \tag{1}$$

The same equation can be rigorously derived from Maxwell's stress tensor approach [13]. For passive media, $1+R>T$, and hence the total pressure is always directed in the direction of beam propagation. Also, for a given reflection, reducing the transmission through dissipation leads to an increase of the pushing force. In the blackbody ($R=0$, $T=0$) and perfect reflection ($R=1$, $T=0$) approximations this formula gives $F = N_i\hbar k$ and $F = 2N_i\hbar k$.

It can be rigorously shown that optical pulling force is forbidden not only in the particular case outlined above but also for a passive object of any shape or composition in a paraxial field [14]. Although different approaches to get around this fundamental restriction have been suggested [15], all of them rely on structuring the incident field [14,16–21], utilizing gain media [22–24], or on modifying the surroundings of the manipulated object [25–28], and hence require either complex techniques or operate at certain conditions. In this work, we revisit the issue of optical forces by stepping out to the complex frequency plane ($\omega = \omega' + i\omega''$) and considering its dynamics upon Fourier untransformable excitations. We show that tailoring of the time evaluation of the light excitation field allows either *enhancement of the repulsive force* or *achieving pulling force for a passive resonator of arbitrary shape and composition*. This unusual response is caused by the "*virtual absorption*" and "*virtual gain*" effects described in the following.

## Results and discussion

To begin with, let us consider the 2-port system and assume the incoming signal to be exponentially growing [ $E_i(t) \propto E_0 \exp(\omega''t)\exp(i\omega't)$ ], Figure 1(b). In this case, as the incoming signal grows in time faster than the decay rate of the cavity, the incoming signal $|E_i|$ and the momentum flux density ($c\varepsilon_0 \hbar k |E_i|^2 / 2\hbar\omega$) can be greater than the corresponding outgoing signal ($|E_t|+|E_r|$) and momentum $[c\varepsilon_0 \hbar k(|E_t|^2 +|E_r|^2)/2\hbar\omega]$, even in the lossless case. The remaining part gets stored in the cavity (virtual loss) until the incoming signal keeps the exponential growth. In this case $1-T(\omega',\omega'')-R(\omega',\omega'') = A > 0$, and the optical pressure can be increased. Here, $A$ denotes the stored energy in the object (cavity). Note that the idea of virtual absorption has been suggested in Ref. [29] for the perfect energy capturing in a resonant cavity for a long time and releasing on-demand. This effect has been experimentally realized in Ref. [30] for elastodynamic waves. In the present work, we show that the virtual absorption effect gives a real contribution to the optical radiation pressure and can be utilized for all-optical manipulation.

Further, if the incoming signal exponentially decays faster than the cavity decay rate, the output signal $|E_t|+|E_r|$ [along with the momentum flux density ($c\varepsilon_0 \hbar k(|E_t|^2 +|E_r|^2)/2\hbar\omega$)] can be *greater than the input one*, Figure 1(c). In this case, the stored energy becomes effectively "negative", $1-T-R = A < 0$, as one would have in a system with a real gain. We refer to this effect as "virtual gain", prove it in the following with analytical and numerical calculations, and demonstrate how it can give rise to negative optical pressure. Note that this complex excitation with negative imaginary frequency has been utilized for obtaining an improved resolution in flat imaging devices [31].

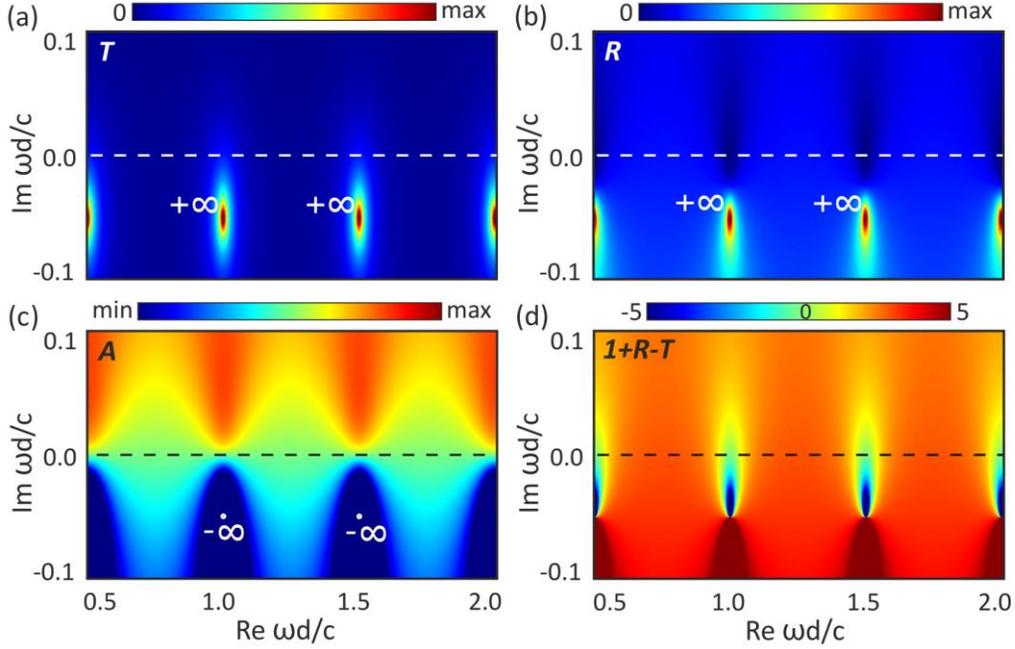

FIG. 2. (a), (b) Transmission and reflection in the complex plane. The dielectric slab is assumed to be lossless, with permittivity $\varepsilon = 40$. (c) Effective absorption in the complex frequency plane calculated as $A(\omega',\omega'') = 1 - R(\omega',\omega'') - T(\omega',\omega'')$. This quantity vanishes at the real axis [ $A(\omega',0) = 0$] due to system losslessness. In the upper (lower) plane, $A(\omega',\omega'')$ is positive (negative) that gives rise to the effective loss (gain) effect. (d) Optical pressure in the complex frequency plane normalized to its value in the perfect absorption case, $F_0 = N_i \hbar k$. As the complex frequency grows in the upper plane, the system can come into the zero transmission regime; as a result, the force reaches $F / F_0 \sim 2$, the value we expect for perfect reflectors. In the lower plane, the difference between the poles of $R$ and $T$ gives the enlarged positive (repulsion) and negative (attraction) force.

As a first example, we consider the case of a dielectric slab with permittivity $\varepsilon = 40$, Figure 2. The material is assumed to be dispersionless and all results are presented in dimensionless frequencies to make the discussion independent of the frequency range. The actual thickness of the slab in our calculations was 500 nm. The transmission and reflection coefficients in the complex plane, Figures 2(a,b), have nontrivial dependence with poles in the lower complex plane. Such point-like exceptional points is an immutable attribute of any resonant structures and associated with the modes of the structure [32]. Interestingly, knowledge of these poles in the complex plane allows retrieving all (linear) electromagnetic properties of the structure *via* the Weierstrass theorem [33]. The reflection coefficient also possesses zeros, associated with the tunneling effect [$R = 0$, $T = 1$ at the real axis] at the Fabry-Perot resonances.

The results of the calculation of the virtual absorption parameter $A(\omega',\omega'') = 1 - R(\omega',\omega'') - T(\omega',\omega'')$ in the complex frequency plane are presented in Figure 2(c). This quantity vanishes at the real axis [$A(\omega',0) = 0$] due to system losslessness. In the upper (lower) plane, virtual absorption $A(\omega',\omega'')$ is positive (negative) that gives rise to the effective loss (gain) effect. This result is fair as long as the excitation has an exponentially increasing or exponentially declining character. Note also that due to the quasi-monochromatic character of the exponential excitation, the calculation results for $A(\omega',\omega'')$, $R(\omega',\omega'')$, and $T(\omega',\omega'')$ are *fair for any time instant*.

The corresponding results of the calculation of the optical pressure are presented in Figure 2(d). Firstly, we note that in the upper plane, which corresponds to the exponentially growing excitation, $E_i(t) \propto E_0 \exp(|\omega''|t)$, as the complex frequency gets increased the system can come into the zero transmission regime [$T = 0$, Figure 2(a,b)]; as a result, the force reaches $F = 2N_i\hbar k$, i.e., the expected value for the perfect reflectors. More remarkably, in the lower plane, which corresponds to the exponentially attenuating excitation, $E_i(t) \propto E_0 \exp(-|\omega''|t)$, the optical pressure experiences the abrupt transition from large positive to large negative values at a fixed real frequency, where the system is expected to experience pulling action from the laser beam, Figure 2(d). This result can be explained as follows. Reflection and transmission coefficients contribute to the resulting radiation force with the opposite sign, and one can expect that their poles compensate each other. However, the presence of reflection zeros in close proximity to the corresponding poles "deform" the poles, and as a result, the optical pressure in the lower complex frequency plane has a nontrivial character with increased and decreased values.

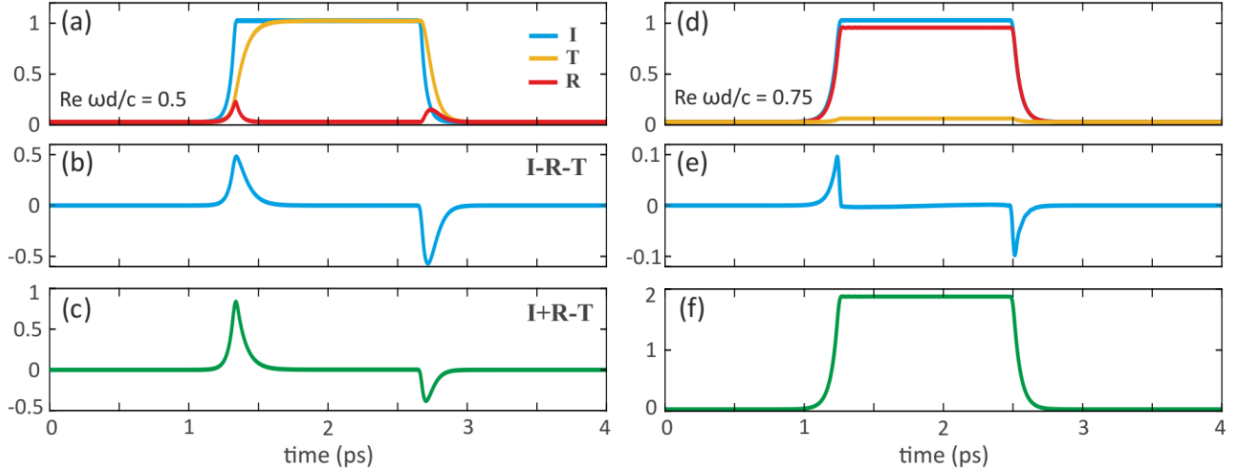

FIG. 3. Transient behavior of incident ($I$, blue curve), transmitted ($T$, yellow curve), and reflected ($R$, red curve) signals obtained in the FDTD simulations and corresponding to (a) the Fabry-Perot resonance [$\text{Re}(\omega d/c) = 0.5$] and (d) the middle between Fabry-Perot resonances [$\text{Re}(\omega d/c) = 0.75$]. During the first time period ($0 \div 1.3$ ps), the oscillation amplitude of the incident signal is increasing exponentially with $\text{Im}(\omega d/c) = 0.05$, then it reaches monochromatic excitation with the same real frequency ($1.3 \div \sim 2.5$ ps), and then it decays exponentially with $\text{Im}(\omega d/c) = -0.05$. The reflection and transmission signals at the monochromatic region correspond to the expected values: ($T \sim 1$, $R \sim 0$) for the Fabry-Perot resonance (a), and ($T \sim 0$, $R \sim 1$) in the intermediate case (d). (b), (e) Transient behavior of the virtual absorption/gain coefficient $A(\omega',\omega'') = 1 - R(\omega',\omega'') - T(\omega',\omega'')$. This value is nonzero only at the exponential pumping and exponential attenuation. (c), (f) Optical radiative force in transient for the chosen excitation. At the Fabry-Perot resonance, the force changes its sign, whereas in the intermediate case it is always positive, in full agreement with Figure 2(d).

To investigate the effect of complex excitation in transient we use FDTD simulations in CST Microwave Studio. To this end, we tailor two incident signals ($I$, blue curve) corresponding to the Fabry-Perot resonance [$\text{Re}(\omega d/c) = 0.5$] and the middle frequency between two Fabry-Perot resonances [$\text{Re}(\omega d/c) = 0.75$], Figure 3(a,d). During the first period ($0 \div 1.3$ ps), the oscillation amplitude of both incident signals is increasing exponentially with $\text{Im}(\omega d/c) = 0.05$, then it reaches monochromatic excitation with the same real frequency ($1.3 \div \sim 2.5$ ps), and then it decays exponentially with $\text{Im}(\omega d/c) = -0.05$. Figures 3(a,d) show the calculated transmitted ($T$, yellow curve), and reflected ($R$, red curve) signals. We note that the reflection and transmission signals at the monochromatic region correspond to the expected values: ($T \sim 1$, $R \sim 0$) for the Fabry-Perot resonance (a), and ($T \sim 0$, $R \sim 1$) in the intermediate case (d), which means good incident pulse quality and the absence of energy transfer to other frequencies. The virtual absorption/gain coefficient is nonzero only at the exponential excitation and exponential attenuation periods as expected for the lossless structure, Figures 3(b, e). The results of the numerical calculation of optical radiative force in transient for the chosen excitation are shown in Figures 3(c, f). We see that at the Fabry-Perot resonance, the force changes its sign, whereas in the intermediate case it always stays positive, in full agreement with analytical results, Figure 2(d).

Thus, these results demonstrate the possibility of getting around the restriction caused by the momentum conservation law and achieving negative optical pressure for exponentially decaying signals. We note that the required decay rate depends on the position of the reflection poles and can be made arbitrarily small with an appropriate chose of mode with a large Q-factor. For instance, the recently introduced concept of optical bound states in the continuum (BICs) [34,35] supporting unboundedly large Q-factor (the pole is unboundedly close to the real axis), would be a promising platform for negative optical forces in paraxial beams slowly decaying in time.

As another example, important from the application viewpoint, we consider the case of a high-index dielectric nanoparticle. Recently, these particles have attracted a lot of interest from researches across many interdisciplinary fields, including quantum optics, nonlinear, and biosensing. For biological applications, these subwavelength dielectric particles are demonstrated to be a fruitful platform for *intracellular* spectroscopy and microscopy [36–38]. In our calculations, we have chosen permittivity $\varepsilon = 16$, which corresponds to c-Si in the visible, Ge in near-IR, and SiC in mid-IR [39].

The scattering cross-section ($Q_{sca}$) of the dielectric nanoparticle in the complex frequency plane is presented in Figure 4(a). As in the previous example, we observe several poles in the lover complex plane that give rise to the corresponding resonances at the real axis [37,40], Figure 4(b). The fundamental resonance is magnetic dipole (MD), whereas the resonance with the largest Q-factor is magnetic quadrupole (MQ). As the frequency grows, the higher-order resonant modes manifest themselves. Here, the anapole state, which corresponds to $Q_{sca} \approx 0$ (it is not exactly zero because of MQ mode) regime is also shown. Recently, the optical radiation force for such a dielectric particle in the monochromatic excitation laser field has been investigated theoretically [41,42] and experimentally [43]. The enhancement of the force around the resonances and the reducing of it at the zero-backscattering Kerker condition [37] have been reported. Although the former is supposed to be used for optical force enhancement, the latter is suggested for stabilization in an optical trap [41]. In Ref. [44] the optical force acting on the Si particles has been utilized for 2D trapping over a substrate and printing onto the substrate by means of radiation pressure. However, the optical pressure in these works is *reported to be always positive for the paraxial optical field* because any act of scattering of the incident photons by the nanoparticle can reduce their forward momentum or, at best, leave it unchanged. In further consideration, we revisit this conclusion and demonstrate how the negative optical radiative force can be achieved in the complex excitation approach.

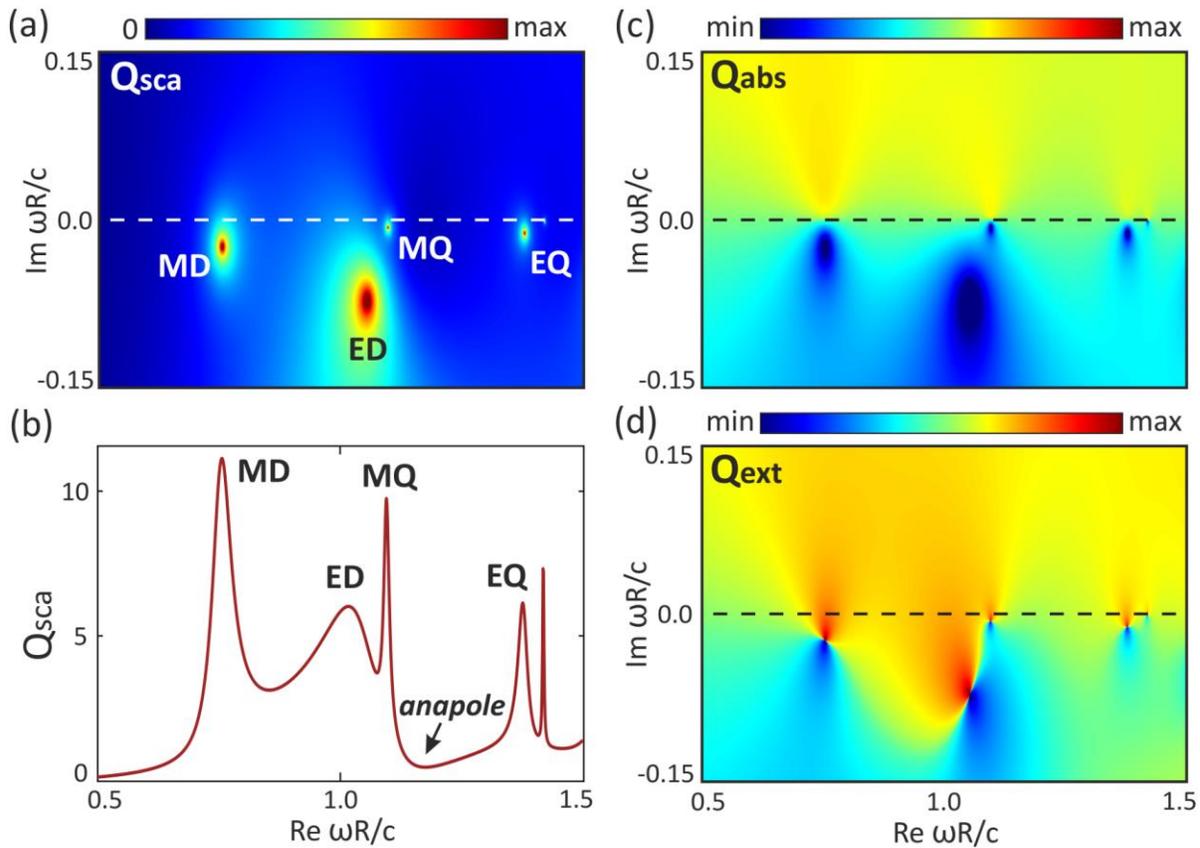

FIG. 4. (a) Scattering cross-section ($Q_{sca}$) of a dielectric particle with permittivity $\varepsilon=16$ in the complex frequency plane. It demonstrates several poles in the lover complex plane that give rise to the corresponding resonances (b) at the real axis (MD, ED, MQ, EQ stand for magnetic dipole, electric dipole, magnetic quadrupole, and electric quadrupole modes). The anapole state corresponds to $Q_{sca} \approx 0$ regime. (c) Absorption cross-section ($Q_{abs}$) and (d) extinction cross-

section ($Q_{ext}$) of the nanoparticle in the complex frequency plane. Despite that the particle is assumed to be lossless [$Q_{abs}(\omega',0) = 0$], the results show finite virtual absorption in the upper plane and negative absorption (gain) in the lower plane.

The radiative optical force in this case of a spherical particle can be calculated analytically by using the method rigorously derived based on the time-averaged Maxwell stress tensor [14,42,45] and found to be consistent with experimental results [43]. According to this method, the time-averaged radiative optical force equals

$$F = \frac{\pi r^2 I}{c}[Q_{ext} - Q_{sca} <\cos\theta>], \qquad (2)$$

$$Q_{sca}<\cos\theta> = \frac{4}{(kR)^2}\sum_l \frac{l(l+2)}{l+1}\mathrm{Re}\left(a_l a_{l+1}^* + b_l b_{l+1}^*\right) + \frac{4}{(kR)^2}\sum_l \frac{2l+1}{l(l+1)}\mathrm{Re}\left(a_l b_l^*\right), \qquad (3)$$

where $Q_{ext}$ is the extinction cross-section ($Q_{ext} = Q_{abs} + Q_{sca}$), $Q_{abs}$ is the absorption cross-section, $l$ is the multipole order, $R$ is the particle radius, $a_l$ and $b_l$ are electric and magnetic Mie scattering amplitudes [45]. To analytically calculate the values of $Q_{ext}$ and $Q_{abs}$ in the complex plane we used the Mie theory [45]. This approach has also been utilized for the analysis of optical forces in spherical particles of active media and can be generalized to nonparaxial excitation fields [16]. In the case of complex excitation, as in the previous consideration of the Fabry-Perot cavity, $Q_{abs}$ has a finite value despite the fact that the particle is lossless [$\mathrm{Im}(\varepsilon) = 0$], Figure 4(c). This stems from the virtual absorption and virtual gain effects discussed above. Physically this virtual loss effect corresponds to the stored energy, whereas the virtual gain effect is caused by the amplitude of energy leaking out from the particle exceeding the amplitude of the incoming energy. Since the extinction cross-section equals the sum of both parts of losses ($Q_{ext} = Q_{abs} + Q_{sca}$), it also demonstrates the alternating behavior in the lower plane, Figure 4(d).

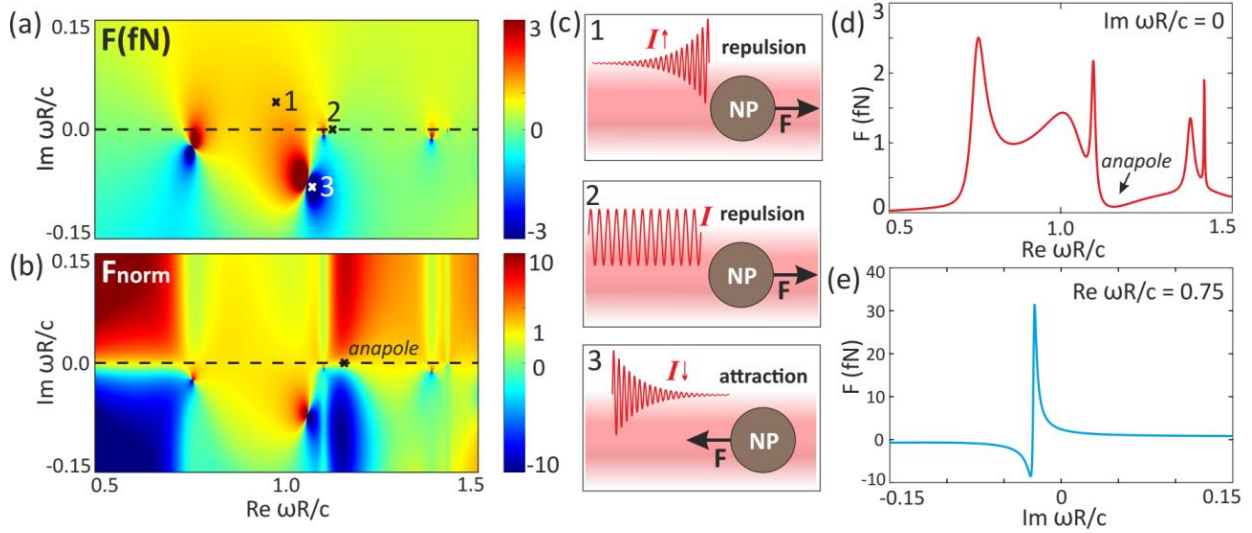

FIG. 5. (a) Optical radiation pressure for the dielectric particle with permittivity $\varepsilon = 16$ in the complex frequency plane. The light intensity is $I = 1$ W/mm$^2$. (b) The same but normalized by its value at the real axis. (c) Scatches of the corresponding regimes denoted by crosses in (a). (d) The value of optical radiation pressure at the real frequency axis [$\text{Im}(\omega R/c) = 0$]. (e) Optical radiation pressure as a function of imaginary frequency $\text{Im}(\omega R/c)$ at the fixed real frequency [$\text{Re}(\omega R/c) = 0.75$], which corresponds to the MD mode.

The calculation results of the optical radiation pressure for the dielectric particle with permittivity $\varepsilon = 16$ in the complex frequency plane are presented in Figure 5(a). Here, by points "1", "2", and "3" we denote the virtual repulsive, repulsive, and virtual pulling regimes, Figure 5(c). The optical force in the complex frequency plane normalized by its value at the real axis is presented in Figure 5(b). For comparison, Figure 5(d) gives the amount of optical radiation pressure at the real frequency axis [$\text{Im}(\omega R/c) = 0$]. This result coincides with the reported theoretical works [41] and existing experimental results [43]. Figure 5(e) demonstrates the optical radiation pressure as a function of imaginary frequency $\text{Im}(\omega R/c)$ at the fixed real frequency [$\text{Re}(\omega R/c) = 0.75$], which corresponds to the MD mode. We observe the characteristic Fano-like transition from enhanced negative to enhanced positive values.

Concluding this discussion, we discuss possible experimental approaches to achieve the reported effects. First of all, as we mentioned above, the rate of exponential grows or decay required to archive enhanced positive or negative radiation force depends on the position of the poles in the complex plane. These poles may lie closely to the real frequency axis in high-Q cavities. An ultimate approach in this way is so-called bound states in the continuum or embedded eigenstates [46–48]. In contrast to conventional optical resonances (e.g., plasmonic, Mie, whispering gallery modes), these states are uncoupled from the continuum of radiative modes, and hence in a lossless scenario, their poles lie on the real frequency axis enabling negative optical pressure for slowly decaying fields. Next, the reported results of this work are rather general and remain fair in the microwave, THz, and optics spectral ranges. In microwaves, the generation of such pulses is rather established experimental technique. Finally, the exponentially decaying fields are "naturally" arise after abrupt turning-off of the excitation of a resonator, which starts releasing its stored energy exponentially with the decay rate of a particular resonant mode. An object situated in its vicinity is expected to manifest the negative optical radiative force.

## Conclusions

In this work, we have revisited the issue of optical forces by stepping out to the complex frequency plane and considering its dynamics upon complex excitations. We have shown that tailoring of the time evaluation of the light excitation field allows either enhancement of the repulsive force or achieving pulling force for a passive resonator of arbitrary shape and composition. We have demonstrated how these effects are linked to virtual gain and virtual loss effects. Virtual gain can be achieved when an appropriate transient decay of the excitation signal makes it weaker than the outgoing signal that carries away greater energy and momentum flux density. In its turn, the virtual loss effect is achieved when the incoming signal exponentially grows in time. The approach has been implemented for the Fabry-Perot cavity and a high refractive index dielectric nanoparticle, a fruitful platform for intracellular spectroscopy and lab-on-a-chip technologies where the proposed technique may found unprecedented capabilities.

## Acknowledgments

The authors thank Prof. Andrea Alú for fruitful discussions.